# TRANSPORT PROTOCOLS FOR LARGE BANDWIDTH-DELAY PRODUCT NETWORKS
## *TCP Extensions and Alternative Transport Protocols*

Rui Policarpo Duarte[1] [2]
*rui.policarpo@tagus.ist.utl.pt*
[1]*Instituto Superior Técnico (IST),*
*Departamento de Engenharia Electrónica e Computadores (DEEC), Lisboa, Portugal*
[2]*Centro de Estudos e Desenvolvimento de Electrónica e Telecomunicações (CEDET),*
*Instituto Superior de Engenharia de Lisboa (ISEL), Lisboa, Portugal*

Abstract: TCP's poor performance is identified as the bottleneck of high-speed networks. Extensions to TCP have been proposed and implemented. Some authors abandon TCP at all and suggest new transport protocols to overcome TCP limitations, at the expense of compatibility. This paper reports a research on the most significant TCP extensions and transport alternatives, and comparison between them. The majority of the solutions pointed out are difficult to compare because they are tailored to specific configurations. Still there is no specific criteria to evaluate performance metrics and comparison is done on the most evident issues.

## 1 INTRODUCTION

Applications using protocols such as HTTP [9], FTP [22], SMTP [21] and even HTCPCP [19] rely on TCP [20] as its transport protocol. It happens that congestion control mechanism in TCP is responsible for under-usage of the available bandwidth. It is the same mechanism that prevents communication stall and permitted networks bandwidth to be scaled thousands of times.

This work points out a known problem, [3, 7], but still there is no unified solution to it. The problem stated in this paper is the large bandwidth-delay product (BDP) in TCP connections as a result of increase of bandwidth in most networks. Compatibility with prior TCP implementations or equipment sets a barrier against new ideas to innovate. Several proposals of congestion control algorithms and reviews to those proposals have been made, e.g. [18]. Focusing on a limited set of key points, they deliver narrow or restrictive benefits to applications. One particular algorithm with optimal performance in one scenario may perform worst than standard TCP in other scenarios. It is frequent to see evaluation of algorithms using different test conditions, so comparison is not straight forward. There is a lack of a generic test-bench to standardize evaluation of contributions.

This work presents the most significant performance metrics that should be taken into account in future developments.

## 2 TCP VARIATIONS

Transmission Control Protocol (TCP) [20] has been one of the cornerstones of the Internet Protocol Suite.

The congestion control algorithm in standard TCP was developed in 1997 [27]. It defines the algorithms for Slow Start, Congestion Avoidance, Fast Retransmit, and Fast Recovery. Several proposals for TCP enhancement have been made since the beginning, e.g., [15, 18]. This standard congestion control mechanism has performed well and it has prevented severe congestion as the Internet magnitude in size, speed, load, and connectivity increased. However, the BDP continues to grow and TCP became a performance bottleneck itself. According to [5], the following four difficulties contribute to the poor performance of TCP in networks with large BDP:

**Packet level:** slow linear packet increase, by one packet per round trip time (RTT), and multiplicative decrease per loss event is too harsh.

**Packet level:** oscillation in congestion window is unavoidable because TCP uses a binary congestion signal (packet loss).

- **Flow level:** maintaining large average congestion windows requires an extremely small equilibrium loss probability.
- **Flow level:** the dynamics is unstable, leading to severe oscillations that can only be reduced by the accurate estimation of packet loss probability and a stable design of the flow dynamics. Delay-based congestion control has been proposed, e.g., in [4, 16, 31].

RFC 2581 [18] explicitly allows certain modifications of these algorithms, including modifications that use the TCP Selective Acknowledgement (SACK) option, and modifications that respond to "partial acknowledgments" (ACKs which cover new data, but not all the data outstanding when loss was detected) in the absence of SACK. TCP NewReno is defined in [11] and introduced modifications to TCP's Fast Recovery algorithm.

Several proposals to overcome the limitation of networks with high BDP have been published. They can be classified into five categories: 1) loss based, 2) delay based, 3) loss & delay based, 4) explicit congestion notification and 5) split connections.

## 2.1 TCP Vegas

TCP Vegas [4] was presented in 1994 as a new TCP implementation that achieved between 40% and 70% better throughput and one-fifth to one-half the losses, when compared with standard TCP. All changes made by TCP Vegas to the standard are confined to the sending side. TCP Vegas proposed some innovations in retransmission mechanism, congestion avoidance mechanism and slow-start mechanism. The congestion avoidance mechanism of TCP Vegas measures the amount of extra data that a connection has in transit. It means the more congestion there is in the network, implies that the sending rate should be reduced. TCP Vegas tries to find a connection's available bandwidth that does not incur this kind of loss. To be able to detect and avoid congestion during slow-start, TCP Vegas allows exponential growth only every other RTT. It happens that when competing with other flows TCP Vegas senses congestion and it draws back freeing bandwidth to be occupied by other standard TCP "bandwidth greedy" flows. It would only be fair if and only if every single machine in the network were using TCP Vegas.

## 2.2 HighSpeed TCP

HighSpeed TCP (HSTCP) was proposed in 2003 by [10], and it was designed to keep congestion metrics within acceptable values, while using large congestion windows. The congestion avoidance mechanism used is a modification of the basic Additional Increase Multiple Decease (AIMD) behavior, dependent on current window size. HSTCP is a modification to TCP's current congestion control mechanisms for high speed links. The current implementations of TCP are limited by the way it controls its congestion windows in the face of losses which result in major disruptions in goodput. One issue with the change to the AIMD algorithms is that HSTCP may impose a certain degree with unfairness as it does not reduce its transfer rate at much as Standard TCP. Similarly, under congestion control, its slow start can be more aggressive. This is especially pronounced given higher loss rates as HighSpeed TCP.

## 2.3 Scalable TCP

Very similar to High Speed TCP, Scalable TCP (STCP) was also defined in [10]. Different additional and multiplying factors are used, in order to keep fairness and have a friendly behavior towards standard TCP. It's goal is to allow high utilization of the network regardless of the under laying technology. It achieved this by implementing the ability to recover from congestion at any `cwnd` size in the same amount of time. STCP was implemented in the network stack of the Linux 2.4.19 operating system. Scalable TCP patch adds the congestion window algorithm changes, scalings to kernel buffers, the removal of special case small packet handling and debug counters.

## 2.4 BIC/CUBIC TCP

Since Linux 2.6.13, BIC [12] had been included in standard Linux distributions and set to the default TCP. Currently, the successor of the BIC, CUBIC [32], is set to default, for kernel versions greater than 2.6.13. The main feature of BIC is its window growth function. Figure below shows the growth function of BIC.

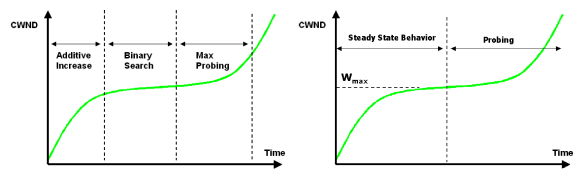

Figure 1: BIC and CUBIC window growth.

A new window growth function gave existence to CUBIC. It has three new features: New window growth function (cubic); New TCP friendly mode; Low utilization detection.

In this new release of BIC, authors introduce a new window growth function - a cubic function. The cubic function shape is similar to the BIC window curve. The function grows much slower than binary increase, which is the logarithmic, near the origin.

CUBIC is instead less aggressive at startup avoiding the additive increase, it does not perform the binary search phase and it soften the max probe phase. Fairness, scalability and friendliness are reported to be good and better than BIC. To further enhance the fairness and stability, CUBIC also clamp the window increment to be no more than Smax per second.

## 2.5 Fast TCP

Fast TCP [5] was proposed in 2003 as a new TCP congestion control algorithm for high-speed long-latency networks. It is a descendent of TCP Vegas. They differ in the way in which the rate is adjusted when the number of packets stored is too small or large. TCP Vegas makes fixed size adjustments to the rate, independent of how far the current rate is from the target rate. FAST TCP makes larger steps when the system is further from equilibrium and smaller steps near equilibrium. Improving speed of convergence and stability.

## 2.6 TCP Peach

This congestion control scheme is an end-to-end solution to improve the throughput performance in satellite networks. TCP-Peach [13] was introduced in 2001. It is based on the use of dummy packets, which are low-priority segments that do not carry any new data to the receiver. TCP-Peach requires the routers along the connection to implement some priority mechanism at the IP layer. TCP Peach implements Congestion Avoidance and Fast Retransmit (the same as TCP Reno), and two new algorithms: Rapid Recovery and Sudden Start, which are based on the use of dummy segments, low priority segments generated by the sender as a copy of the last transmitted data packet. The main features of TCP-Peach is that it only requires modifications in the end user behaviors and that it is compatible with traditional TCP implementations.

## 2.7 TCP Hybla

TCP Hybla [6]tries to cope with the disadvantage of TCP connections that incorporate a terrestrial or satellite radio link with respect to entirely wired connections, because of their longer round trip times (RTTs). It is the resultant of an analytical evaluation of the congestion window dynamics in the TCP standard versions (Tahoe, Reno, NewReno). TCP Hybla approach suggests the necessary modifications to remove the performance dependence on RTT, it reduces the severe penalization suffered by wireless, and especially satellite, TCP connections. It does not infringe the end to end semantics of TCP and is compatible with other TCP enhancements.

## 2.8 TCP Westwood

TCP Westwood (TCPW) [24], is a sender-side-only modification to TCP NewReno that is intended to better handle large BDP paths (large pipes), with potential packet loss due to transmission or other errors (leaky pipes), and with dynamic load (dynamic pipes). TCP Westwood relies on mining the ACK stream for information to help it better set the congestion control parameters: Slow Start Threshold (`ssthresh`), and Congestion Window (`cwin`). In TCP Westwood, an "Eligible Rate" is estimated and used by the sender to update `ssthresh` and `cwin` upon loss indication, or during its "Agile Probing" phase, a proposed modification to the well-known Slow Start phase. The resultant performance gains in efficiency, without undue sacrifice of fairness, friendliness, and stability have been reported in numerous papers that can be found on this web site. TCP Westwood bandwidth estimation algorithm does not work well in the presence of reverse traffic due to `ack` compression.

## 2.9 TCP Low Priority

TCP Low Priority (TCP-LP) [17] presents an approach to the problem of service prioritization among different traffic classes in the Internet. TCP-LP algorithm uses only the excess network bandwidth as compared to the "fair share" of bandwidth as targeted by TCP. TCP-LP congestion control mechanisms are the use of one-way packet delays for early congestion indications and a TCP-transparent congestion avoidance policy. The existence of a mechanism for early congestion indication via inferences of one-way packet delays. The policy of congestion avoidance is sensitive to early congestion indicators. The feasibility of TCP-transparent congestion control, under heterogeneous round trip times, is proved with a queuing model.

## 2.10 TCP Real

The idea behind TCP-Real [33] is to combine into TCP the concept of "wave" from Wave-and-Wait protocol (WWP)[30] without changing the semantics of TCP and without violating the established standards of Additive-Increase/Multiplicative-Decrease-based congestion control. Every RTT, it sends a wave, a pattern of packets side by side, known by the receiver to permit determining the level of congestion. TCP-Real is based on the accuracy of congestion level estimation. It monitors the level of congestion by measuring the data-receiving rate. The sender adjusts the sending window before the packet loss occurs. Both the sender and the receiver are aware of the current window size (wave level), since the congestion window is now included in the header.

## 2.11 I-TCP

I-TCP [1] splits the transport link at the wireline-wireless border. The base station maintains two TCP connections, one over the fixed network, and another over the wireless link. This way, the poor quality of the wireless link is hidden from the fixed network. By splitting the transport link, I-TCP does not maintain end-to-end TCP semantics, i.e. I-TCP relies on the application layer to ensure reliability. A split connection TCP implementation terminates the TCP connection received from an end system and establishes a corresponding TCP connection to the other end system. In a distributed Performance Enhancement Proxy (PEP) [14] implementation, this is typically done to allow the use of a third connection between two PEPs optimized for the link. This might be a TCP connection optimized for the link or it might be another protocol, for example, a proprietary protocol running on top of UDP.

## 3 Transport Alternatives

Other transport protocols having different applicability contexts have been proposed besides TCP. In the following sections there's a brief overview of the most relevant ones.

### 3.1 Wave and Wait

The Wave and Wait Protocol (WWP) [30] attempts to keep the amount of data transmitted below perceived network congestion. The higher the detected congestion imminence-level in the network the less it attempts to transmit, minimizing the need for duplicate data retransmission due to congested routers losing packets and so on.

Connection is first established using a six-way handshake. This six-way exchange is also used to determine current congestion conditions in the network. A sender sends a "wave" to the receiver consisting of a number of fixed-sized data segments, and then waits for a response. The number of segments in a wave is set according to the current "wave level" which is determined by the receiver in line with the estimated prevailing congestion level, and is communicated to the sender. The less the perceived congestion risk in the network, the higher the "wave level" and the more segments that a wave comprises.

### 3.2 XCP

eXplicit Control Protocol (XCP) [8], generalizes the Explicit Congestion Notification proposal (ECN) [23]. In addition, XCP introduces the new concept of decoupling utilization control from fairness control. Using a control theory framework, XCP is stable regardless of the link capacity, the round trip delay, and the number of sources. Extensive packet-level simulations show that XCP outperforms TCP in both conventional and high BDP environments. Further, XCP achieves fair bandwidth allocation, high utilization, small standing queue size, and near-zero packet drops, with both steady and highly varying traffic.

### 3.3 XTP

Xpress Transport Protocol (XTP) [25] is a transport layer protocol for high-speed networks developed to replace TCP. XTP provides protocol options for error control, flow control, and transmission rate and burst size control. Instead of separating protocols for each type of communication, XTP controls packet exchange options to produce different models, e.g. reliable datagrams, transactions, unreliable streams, and reliable multicast connections. XTP provides mechanisms for shaping rate control and flow control independently, it does not define congestion avoidance algorithms and flow control operates on end-to-end buffer space. The protocol functions incorporated in XTP are structured around four explicit principles: separation of paradigm from policy, separation of rate and flow control, explicit support for multicast, and data delivery service independence.

### 3.4 WTCP

In WTCP protocol [26] the base station plays a pivotal role by buffering the TCP segments and locally retransmitting them based on timeouts as well as duplicate acknowledgments. WTCP uses flow control for the wireless link, and maintains end-to-end TCP semantics. Furthermore, WTCP hides the time spent by the base station for local recovery so that TCP's round trip time estimation at the source is not affected. This is critical since otherwise the ability of the source to effectively detect congestion (i.e. losses due to buffer overflow) in the wireline network can be impaired.

## 3.5 SCTP

Stream Control Transmission Protocol (SCTP) is described in RFC2960 [28], and updated by RFC3309 [29] and it is viewed as a layer between the SCTP user application and a connectionless packet network service such as IP. SCTP offers a reliable message-based data transfer service with the following characteristics: TCP-like congestions control, support for multi-homed hosts, more robustness against denial-of-service attacks, and support for optional out-of-order data delivery. SCTP's congestion control and loss recovery mechanisms closely resemble those of TCP SACK with NewReno implementation described above, with the addition of congestion window growth due to byte counting, and hence the performance due to congestion control and loss recovery in a satellite environment should be similar.

## 4 PERFORMANCE METRIC AND COMPARISON

Metrics to evaluate the performance of a protocol can be expressed in terms of throughput (in packets/s or bytes/s) and latency. Transport protocols can be compared according to the following considerations:

**Congestion Control** - to avoid communication stall, the protocol must have a congestion control mechanism.

**TCP Friendliness** - the new protocol should be compatible with TCP. Under congestion it doesn't take more bandwidth than standard TCP flows under similar conditions.

**Fairness** - when competing with other flows, from different protocols or the same, it should share link occupation with other flows.

**Flow Model** - it eases protocol implementation and debug. Usually it is done graphically or with some higher level description language.

**Deployment** - some protocols demand modification of the operating system kernel, while others are installed in user level being instantly available.

Table 1 summarizes the algorithms according to their category, introduced in section 2.

| Category | Algorithm |
|---|---|
| Loss based | HighSpeed, Scalable, BIC/CUBIC, Peach, Hybla |
| Delay based | Vegas, Fast, Westwood, LowPriority, Real |
| Loss & delay based | Wave and Wait, XTP |
| Congestion notification | XCP, SCTP |
| Split connections | I-TCP, WTCP |

Table 1: Categorization of the several transport proposals.

## 5 CONCLUSIONS AND FUTURE DEVELOPMENTS

TCP performance may be impacted by factors such as BDP, round-trip time, non-congestion losses, and bandwidth asymmetry. Barakat et al., in [3], state that the two main problems of TCP that remain to be solved are burstiness and the coupling between congestion detection and error control.

Proposed works presented here use different evaluation metrics. Heterogeneity of the traffic flows turns analysis difficult and introduces more complexity to the problem. Recently a group suggested a common basic suite to evaluate TCP performance [2].

Despite the broad spectrum of solutions presented, another possibility is to pass information across layers surrounding TCP, where it interacts and its dependence exist.

## 6 ACKNOWLEDGMENTS

The author wishes to thank Prof. José Brázio from Instituto Superior Técnico/Instituto de Telecomunicações for the opportunity to work with him and do research on this specific subject, and also the members of the Networks and Multimedia group at Instituto de Telecomunicações for their support and ideas.